# On the motivation and foundation of Natural Time Analysis: Useful remarks

Panayiotis A. VAROTSOS[1], Nicholas V. SARLIS[1], and Efthimios S. SKORDAS[1]

[1]Department of Solid State Physics and Solid Earth Physics Institute, Faculty of Physics, School of Science, National and Kapodistrian University of Athens, Panepistimiopolis, Zografos 157 84, Athens, Greece; e-mail: pvaro@otenet.gr

**Abstract**

Since its introduction in 2001, natural time analysis has been applied to diverse fields with remarkable results. Its validity has not been doubted by any publication to date. Here, we indicate that frequently asked questions on the motivation and the foundation of natural time analysis are directly answered if one takes into account the following two key points that we have considered as widely accepted when natural time analysis was proposed: First, the aspects on the energy of a system forwarded by Max Planck in his Treatise on Thermodynamics. Second, the theorem on the characteristic functions of probability distributions which Gauss called Ein Schönes Theorem der Wahrscheinlichkeitsrechnung (beautiful theorem of probability calculus). The case of the time series of earthquakes and of the precursory Seismic Electric Signals are discussed as typical examples.

## 1. INTRODUCTION

Since 2001 (Varotsos et al., 2001; 2002a,b), it has been proposed that unique dynamic features hidden behind can be revealed from the time series of complex systems, if we analyze them in terms of a new time domain termed natural time $\chi$. Examples of data analysis in this time domain carried out by a multitude of authors (e.g., Vargas et al., 2015, Flores-Márquez et al., 2014; Ramirez-Rojas et al., 2011, Rundle et al., 2012; Holliday et al., 2006) have appeared in diverse fields, including Biology, Cardiology, Condensed Matter Physics, Environmental Sciences, Geophysics, Physics of Complex Systems, Statistical Physics, and Seismology. Several of these applications have been compiled in a monograph by Varotsos et al. (2011a), where the foundations of natural time analysis have been also explained in detail by providing the necessary mathematical background in each step. It is the objective of this short paper to shed more light on frequently asked questions related to the motivation and the foundation of natural time analysis. For example, although this analysis does not make use of any adjustable parameter, a question raises on why the selection of the normalized energy for each event is preferred in this analysis (see below) compared to other physical quantities. For a time series comprising $N$ events, we define as natural time $\chi_k$ for the occurrence of the $k$-th event the quantity $\chi_k = k/N$. In doing so, we ignore the time intervals between consecutive events, but preserve their order and energy $Q_k$. The analysis in natural time is carried out (Varotsos et al., 2001; 2002a,b) by studying the evolution of the pair ($\chi_k, p_k$), where the quantity

$$p_k = \frac{Q_k}{\sum_{n=1}^{N} Q_n} \qquad (1)$$

is the normalized energy for the $k$-th event, and using the normalized power spectrum (cf. $\omega$ stands for the angular natural frequency):

$$\Pi(\omega) \equiv |\Phi(\omega)|^2 \qquad (2)$$

defined by

$$\Phi(\omega) = \sum_{k=1}^{N} p_k \exp(i\omega\chi_k) \qquad (3)$$

$\Phi(\omega)$ is (Varotsos et al., 2011a) the characteristic function of $p_k$ for all $\omega \in R$ since $p_k$ can be regarded as a probability for the occurrence of the $k$-th event at $\chi_k$. This is obvious for dichotomous signals as it is frequently the case of Seismic Electric Signals (SES) activities (but holds for other signals as well (Varotsos et al., 2001; 2002a,b; 2009; 2011a), see also Section 3), because $Q_k$ is then proportional to the duration of the $k$-th event and $\sum_{n=1}^{N} Q_n$ to the total duration of $N$ events recorded, thus the ratio $Q_k / \sum_{n=1}^{N} Q_n (= p_k)$ gives the probability to observe the $k$-th event among the other events (see Fig. 1) at the natural time $\chi_k = k/N$.

In natural time analysis, the behavior of $\Pi(\omega)$ is studied at $\omega \to 0$, because all the moments of the distribution of $p_k$ can be estimated from the derivatives $d^m \Phi(\omega)/d\omega^m$ (for $m$ positive integer) of the characteristic function $\Phi(\omega)$ at $\omega \to 0$ (see p.512 of Feller (1971)). For this purpose, a quantity $\kappa_1$ was defined (Varotsos et al., 2001; 2002a) from the Taylor expansion (see also the Appendix):

$$\Pi(\omega) = 1 - \kappa_1 \omega^2 + \kappa_2 \omega^4 + \ldots \qquad (4)$$

where

$$\kappa_1 = \langle \chi^2 \rangle - \langle \chi \rangle^2 = \sum_{k=1}^{N} p_k \chi_k^2 - \left( \sum_{k=1}^{N} p_k \chi_k \right)^2 \qquad (5)$$

It has been shown (Varotsos et al., 2011a,b) that $\kappa_1$ becomes equal to 0.070 at the critical state for a variety of dynamical systems. Once $N$ consecutive events have been observed, the $k$-th event that occurred at natural time $\chi_k = k/N \, (k \leq N)$ will be hereafter called, for the sake of convenience, "$\frac{k}{N}$ event". Note that upon the occurrence of an additional event, the value of $\chi_k$ changes from $k/N$ to $k/(N+1)$ together with the change of $p_k$ from $Q_k / \sum_{n=1}^{N} Q_n$ to $Q_k / \sum_{n=1}^{N+1} Q_n$, thus leading to changes of $\Pi(\omega)$ and $\kappa_1$ as well. Such changes are important for example when analyzing in natural time the small earthquakes that occur after the initiation of an SES activity in the candidate epicentral area in order to estimate the occurrence time of the forthcoming mainshock (e.g., Varotsos et al., 2008, see also Huang, 2015). Hence, when dealing with the possible outcomes of a yet-to-be-performed experiment (as well as when studying the behavior of an evolving dynamical system), where the total number of events that will be recorded is not known in advance, the $\chi_k$ values that are always discrete rational numbers in (0,1] vary until the occurrence of the last event.

Using $p(\chi) = \sum_{k=1}^{N} p_k \delta\left( \chi - \frac{k}{N} \right)$, which is the distribution corresponding to $p_k$, the normalized power spectrum $\Pi(\omega)$ of Eq. (2) can be rewritten in terms of $p(\chi)$ as (Varotsos et al., 2011b)

$$\Pi(\omega) = \int_0^1 \int_0^1 p(\chi) p(\chi') \cos[\omega(\chi - \chi')] d\chi d\chi' \qquad (6)$$

and its Taylor expansion around $\omega \to 0$ leads to the value

$$\kappa_1 = \frac{1}{2} \int_0^1 \int_0^1 p(\chi) p(\chi') (\chi - \chi')^2 d\chi d\chi' \qquad (7)$$

Alternative expressions for $\Pi(\omega)$ and $\kappa_1$ are given in the Appendix.

## 2. RANDOM VARIABLE. CHARACTERISTIC FUNCTION. BACKGROUND.

### 2.1 Random variable.

Since 1933, Kolmogorov has first made clear that a random variable is nothing but a measurable function on a probability space. Let (Ω, A, P) be a probability space, where Ω stands for the sample space (the set of world states, sometimes called outcomes), A the event space (the set of subsets of Ω) and P the probability measure. A single-valued real-valued function *X=X(ω)* defined on Ω ($\omega \in \Omega$) is called a random variable if for any real *x* the set $\{\omega : X(\omega) < x\}$ belongs to the class A (e.g., see Random variable. Encyclopedia of Mathematics, available from: http://www.encyclopediaofmath.org/index.php? title=Random_variable&oldid=29510. Accessed: 2015-02-02.). In simple words, a random variable X is a function that associates a unique numerical value with every outcome of an experiment (e.g., see Statistics Glossary v1.1 (STEPS), available from: http://www.stats.gla.ac.uk/steps/glossary/probability_distributions.html#probdistn, Accessed: 2015-02-02.), or more simply (e.g., see Summary of Chapter 8 in Finite Mathematics and Finite Mathematics & Applied Calculus, available from: http://www.zweigmedia.com/ThirdEdSite/Summary7.html, Accessed: 2015-02-02.) it is just a rule that assigns a numerical value to each outcome in the sample space of an experiment. Also a random variable is sometimes described as a variable whose value is subject to variations due to chance, i.e., randomness in a mathematical sense, e.g., see p.391 of Yates et al (2002) (but see also p.500 of Jaynes, 2003, where the following is written: "However, although the property of being "random" is considered a real objective attribute of a variable, orthodoxy has never produced any definition of the term "random variable" that could actually be used in practice to decide whether some specific quantity, such as the number of beans in a can, is or is not "random". Therefore, although the question "which quantities are random? " is crucial to everything an orthodox statistician does, we are unable to explain how he actually decides this; we can only observe what decisions he makes."). A random variable differs essentially from other mathematical variables since it conceptually does not have a single, fixed value (even if unknown), but it can take on a set of possible different values each with an associated probability (if discrete) or a probability density function (if continuous).

A random variable's possible values might represent the possible outcomes of either a past experiment whose already existing value is uncertain (e.g., as a result of incomplete information or imprecise measurements), or a yet-to-be-performed experiment. A typical example is the case of the analysis of a series of seismic events, which for instance should be carried out for the time series of small earthquakes occurring in the candidate area after the SES initiation in order to estimate the occurrence time of an impending mainshock, as mentioned earlier. In this example, the values of $k/N$ depend of course on the magnitude threshold adopted. In addition, for each threshold selected, the values of $k/N$ are subject to variations due to experimental error in the determination of the magnitude, which may result in the observation or not of the smaller events, especially the ones in the vicinity of the threshold in a way explained in Varotsos et al., (1996). In other words, due to the experimental error, events slightly exceeding the threshold selected may not be reported in the measurement while others slightly smaller than the threshold may be reported thus affecting the *k/N* values. Also, we clarify that a random variable's values may conceptually represent either the results of an "objectively" random process (e.g., rolling a die) or the "subjective" randomness that results from incomplete knowledge of a measurable quantity. The latter is the case of the aforementioned example of analyzing a series of seismic events in which, as mentioned, an experimental error in the magnitude determination may affect the values of *k/N*.

## 2.2 Characteristic function.

Let *X* be a random variable. The characteristic function of a continuous distribution with cumulative distribution function *F(x)* is defined as

$$(\omega \in R): \Phi(\omega) = E\left[e^{i\omega X}\right] = \int_{-\infty}^{\infty} e^{i\omega x} dF(x) \qquad (8)$$

which is a complex-valued function. *E* stands for the expectation value. For a discrete distribution on the non-negative integers *j*, it is defined as

$$\Phi(\omega) = E\left[e^{i\omega X}\right] = \sum_{j=0}^{\infty} e^{i\omega j} \Pr[X = j] \qquad (9)$$

where Pr[*X=j*], or simply $p_j$ (if we follow the symbol used in Eq. (1) in Section **1**), denotes for the random variable *X* its associated probability at the non-negative integer value *X=j*.

The characteristic function *uniquely* determines the probability density function of a continuous distribution (see p.509 of Feller (1971); see also p.48 of Johnson et al., (1992)); we have

$$f(x) = \frac{1}{2\pi} \int_{-\infty}^{\infty} e^{-i\omega x} \Phi(\omega) d\omega \qquad (10)$$

Gauss called this theorem Ein Schönes Theorem der Wahrscheinlichkeitsrechnung (beautiful theorem of probability calculus). The corresponding inversion formula for discrete distributions on the non-negative integers is

$$\Pr[X = x] = \frac{1}{2\pi} \int_{-\pi}^{\pi} e^{-i\omega x} \Phi(\omega) d\omega \qquad (11)$$

## 3. CHARACTERISTIC FUNCTION AND NATURAL TIME ANALYSIS.

In accordance to Eq. (9), the function $\Phi(\omega)$ in Eq. (3) introduced in natural time analysis (Varotsos et al., 2001; 2002a,b), i.e.,

$$\Phi(\omega) = \sum_{k=1}^{N} p_k \exp(i\omega \frac{k}{N}) \qquad (12)$$

constitutes the characteristic function of the random variable $(X =) \frac{k}{N}$. This is so, because $p_k$ -as given by Eq.(1) - can be regarded (Varotsos et al., 2001; 2002a,b) as the probability associated with $X = \frac{k}{N}$ (Moreover $\Phi(\omega)$ of Eq.(12) is positive definite for all $\omega \in R$, $\Phi(0)=1$ and the map $\omega \to \Phi(\omega)$ is continuous at the origin, thus satisfying the conditions of the theorem 1.1.12 in p.17 of Applebaum, 2003, to be characteristic function of a probability distribution). In other words, $p_k$ is the probability for observing the $\frac{k}{N}$ event. The latter becomes clear if we focus on the question "What is energy?" by considering the authoritative aspects of Max Planck to which we now turn. These aspects also show an insight into the use of the quantity of energy when natural time analysis was proposed (Varotsos et al., 2001; 2002a,b).

### 3.1 What is energy?

Recent relevant books (e.g., see Coopersmith, 2010) note that "most physicists are unable to provide a satisfactory answer to this question" (see also Crystal, 2011).

Max Planck, in §58 of his Treatise on Thermodynamics (see p.41 in Planck, 1945) states:

"The energy of a body, or system of bodies, is a magnitude depending on the momentary condition of the system. In order to arrive at a definite numerical expression for the energy of a system in a given state, it is necessary to fix upon a certain normal arbitrarily selected state (e.g., 0ºC and atmospheric pressure). The energy of the system in a given state, referred to the arbitrarily selected state, is then equal to the algebraic sum of the mechanical equivalents of all the effects produced outside the system when it passes in any way from the given to the normal state. The energy of a system is, therefore, sometimes briefly denoted as the faculty to produce external effects".

This definition suggests the following answer to the aforementioned question on "what is energy", e.g., for a recent reference see Bauer (2011): *"The energy of the system is a measure of its presence in the universe"*.

We now proceed to the change of energy, $U_1 - U_2$, accompanying the transition of the system from a state 1 to a state 2 (which may probably happens when observing an event). Max Planck in §63 of his Treatise (see pp.44-45 in Planck, 1945) states:

"The energy, as stated, depends on the momentary condition of the system. To find the change of energy, $U_1 - U_2$, accompanying the transition of the system from a state 1 to a state 2, we should, according to the definition of the energy in §58 [this is the definition mentioned above], have to measure $U_1$ as well as $U_2$ by the mechanical equivalent of the external effects produced in passing from the given states to the normal state. But, supposing we so arrange matters that the system passes from state 1, through state 2, into the normal state, it is evident then that $U_1 - U_2$ is simply the mechanical equivalent of the external effects produced in passing from 1 to 2. The decrease of the energy of a system subjected to any change is, then, the mechanical equivalent of the external effects resulting from that change; or, in other words, the *increase of the energy* of a system which undergoes any change, is equal to the mechanical equivalent of the heat absorbed and the work expended in producing the change:

$$U_1 - U_2 = Q + W \qquad (13)$$

$Q$ is the mechanical equivalent of the heat absorbed by the system, e.g. by conduction, and $W$ is the amount of work expended on the system. …"

and in §64 of his Treatise Max Planck clarifies:

"The difference $U_1 - U_2$ may also be regarded as the energy of the system in state 2, referred to state 1 as the normal state …"

Hence, the aforementioned answer to the question on "what is energy?" also holds for the difference $U_1 - U_2$.

**3.2 Why it is logical to consider p_k as a probability for the observation of the *k/N*-event**

In the light of the above authoritative aspects of Max Planck summarized by Bauer (2011), and assuming that $N$ events have been observed, we think along the following lines: Since $Q_k$ stands for the energy of the *k*-th event and $\sum_{n=1}^{N} Q_n$ for the total energy of N events, their ratio

$Q_k / \sum_{n=1}^{N} Q_n (= p_k)$ can be considered as a probability for the occurrence (observation) of the $\frac{k}{N}$-event. (The quantities $p_k$ are of course, probabilities according to Kolmogorov (1956) since $\sum_{k=1}^{N} p_k = 1$.) This, however, should not be considered as being illogical with the following claim: "this is illogical since one never has the situation that the $\frac{k}{N}$ event occurred with probability 20% and did not occur with probability 80%, for example". This should be understood in the context that 80% refers to the sum of the probabilities of all the other *N−1* events to occur.

## 4. CONCLUSIONS.

The analysis in natural time χ is based on the study of the evolution of the pair ($\chi_k, p_k$). On the basis of the authoritative aspects forwarded by Max Planck concerning the energy of a system, it becomes evident that $p_k$ can be considered as the probability for observing the event at $\chi_k = k/N$. By means of $\chi_k$ and $p_k$ that are experimentally accessible, the characteristic function $\Phi(\omega)$ is obtained. This function, upon recalling the theorem which Gauss called "beautiful theorem of probability calculus", uniquely determines the probability distribution of $p_k$. Thus, in short, the transparency of the analysis in natural time cannot be doubted, in any terms, since it is based on celebrated aspects in Physics and in Probability Calculus, and in addition it does not make use of any adjustable parameter.

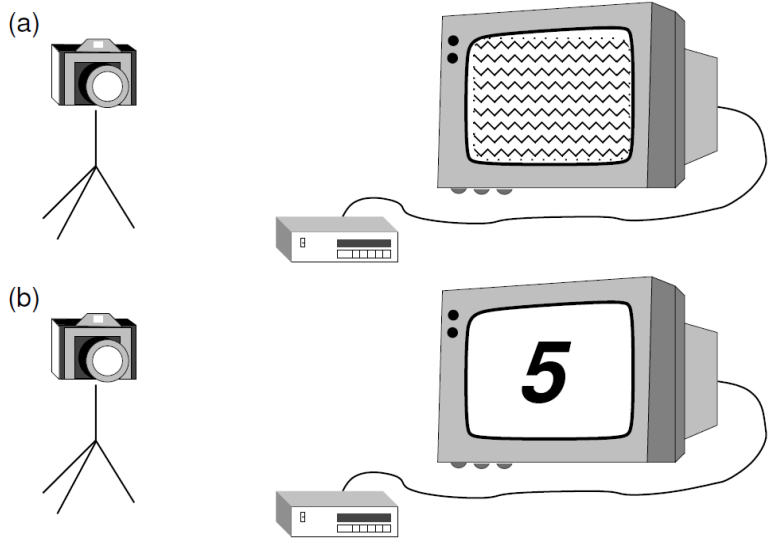

Fig. 1. Assume that $N$ events have been recorded: The observer reports the occurrence of each consecutive event as long as it lasts, e.g., the fifth one (b), but marks nothing during periods of absence of events (a). The reading is replayed and a camera takes a snapshot. Upon discarding the "empty" snapshots, the probability to observe, for example the fifth event, is $Q_5 \Big/ \sum_{n=1}^{N} Q_n$.

# APPENDIX

## ALTERNATIVE EXPRESSIONS FOR $\Pi(\omega)$ AND $\kappa_1$

We first give a general expression for $\Pi(\omega)$ valid for any value of $\omega$:

By writing $\Phi(\omega) = \sum_{k=1}^{N} p_k e^{i\omega \chi_k}$ and $\Phi^*(\omega) = \sum_{m=1}^{N} p_m e^{-i\omega \chi_m}$ we have:

$$\Pi(\omega) = \Phi(\omega)\Phi^*(\omega) = \sum_{k,m=1}^{N} p_k p_m e^{i\omega(\chi_k - \chi_m)} = \sum_{k=1}^{N} p_k^2 + \sum_{k=1}^{N-1} \sum_{i=1}^{N-k} p_k p_{k+i} \underbrace{(e^{i\omega \chi_i} + e^{-i\omega \chi_i})}_{2\cos(\omega \chi_i)}, \quad \text{(A1)}$$

where we used the fact that

$$\chi_{k+i} - \chi_k = \chi_i \qquad (A2)$$

because $\chi_l = l/N$.

Since $\sum_{k=1}^{N} p_k = 1 \Rightarrow \left(\sum_{k=1}^{N} p_k\right)^2 = 1 \Rightarrow \sum_{k=1}^{N} p_k^2 + 2\sum_{k=1}^{N-1}\sum_{i=1}^{N-k} p_k p_{k+i} = 1$, we have:

$$\Pi(\omega) = 1 - 2\sum_{k=1}^{N-1}\sum_{i=1}^{N-k} p_k p_{k+i} + 2\sum_{k=1}^{N-1}\sum_{i=1}^{N-k} p_k p_{k+i} \cos(\omega \chi_i) \qquad (A3)$$

Thus,

$$\Pi(\omega) = 1 - 2\sum_{k=1}^{N-1}\sum_{i=1}^{N-k} p_k p_{k+i}[1 - \cos(\omega \chi_i)] = 1 - 4\sum_{k=1}^{N-1}\sum_{i=1}^{N-k} p_k p_{k+i} \sin^2\left(\frac{\omega \chi_i}{2}\right) \qquad (A4)$$

valid for *any* value of $\omega$.

When $\omega \to 0$ (and since $\max[\chi_i] = 1$), we have $\sin(\omega \chi_i/2) \approx \omega \chi_i/2$ and hence Eq. (A4) simplifies to:

$$\Pi(\omega) \approx 1 - \omega^2 \sum_{k=1}^{N-1}\sum_{i=1}^{N-k} p_k p_{k+i} \chi_i^2 \qquad (A5)$$

By comparing with Eq. (4) and using Eq.(A2), we find for the $\kappa_1$ value the following expression:

$$\kappa_1 = \sum_{k=1}^{N-1}\sum_{i=1}^{N-k} p_k p_{k+i}(\chi_{k+i} - \chi_k)^2 \qquad (A6)$$

Note there exists compatibility of the $\kappa_1$ value obtained from Eq.(7) with Eq.(A6), which can be shown as follows:

We consider that

$$\sum_{l=1}^{N}\sum_{m=1}^{N} p_l p_m (\chi_l - \chi_m)^2 = 2\sum_{k=1}^{N-1}\sum_{i=1}^{N-k} p_k p_{k+i}(\chi_{k+i} - \chi_k)^2 \qquad (A7)$$

In view of Eq.(A7), Eq.(A6) turns to

$$\kappa_1 = \frac{1}{2}\sum_{l=1}^{N}\sum_{m=1}^{N} p_l p_m (\chi_l^2 - 2\chi_l\chi_m + \chi_m^2) =$$

$$= \frac{1}{2}\left\{\sum_{l=1}^{N} p_l \chi_l^2 \underbrace{\sum_{m=1}^{N} p_m}_{1} - 2\sum_{l=1}^{N} p_l \chi_l \sum_{m=1}^{N} p_m \chi_m + \underbrace{\sum_{l=1}^{N} p_l}_{1} \sum_{m=1}^{N} p_m \chi_m^2 \right\}$$

$$= \frac{1}{2}\left\{\langle \chi^2 \rangle - 2\langle \chi \rangle^2 + \langle \chi^2 \rangle\right\} = \langle \chi^2 \rangle - \langle \chi \rangle^2 \qquad (A8)$$

Hence in general

$$\kappa_1 = \langle \chi^2 \rangle - \langle \chi \rangle^2 = \frac{1}{2}\int_0^1\int_0^1 p(\chi)p(\chi')(\chi - \chi')^2 d\chi d\chi', \qquad (A9)$$

i.e., Eq.(7), or

$$\kappa_1 = \langle \chi^2 \rangle - \langle \chi \rangle^2 = \sum_{k=1}^{N-1}\sum_{i=1}^{N-k} p_k p_{k+i}(\chi_{k+i} - \chi_k)^2, \qquad (A10)$$

i.e., Eq.(A6) shown above.

Alternatively, if instead of Eq.(A1), we use the following expression for $\Pi(\omega)$:

$$\Pi(\omega) = \sum_{k=1}^{N} p_k^2 + \sum_{l=1}^{N-1}\sum_{i=1}^{N-l} p_i p_{i+l} \underbrace{(e^{i\omega\chi_l} + e^{-i\omega\chi_l})}_{2\cos(\omega\chi_l)} \qquad (A11)$$

we obtain

$$\Pi(\omega) = 1 - 4 \sum_{l=1}^{N-1} \sum_{i=1}^{N-l} p_i p_{i+l} \sin^2\left(\frac{\omega \chi_l}{2}\right), \tag{A12}$$

which, when $\omega \to 0$, leads to

$$\kappa_1 = \sum_{l=1}^{N-1} \left( \sum_{i=1}^{N-l} p_i p_{i+l} \right) \chi_l^2 . \tag{A13}$$